\documentclass[12pt]{iopart}
\usepackage{iopams}

\newcommand{\la}{\lambda}
\newcommand{\Aa}{\mathcal{A}}
\newcommand{\Bb}{\mathcal{B}}

\newcommand{\z}{\zeta}

\newcommand{\R}{\mathbb R}
\newcommand{\Z}{\mathbb Z}

\def\N{\mathbb{N}}
\def\C{\mathbb{C}}
 
\def\Id{\mathrm{Id}}

\newcommand{\text}{\mbox}
\newcommand{\eqref}[1]{(\ref{#1})}\newcommand{\mathscr}{\rm}

\def\max{\mathrm{max}}

\begin{document}
 \title[Unusual poles and logarithmic singularities of zeta functions]
 {The ubiquitous $\zeta$-function and some of its ``usual" and
``unusual" meromorphic properties}
 \author{Klaus Kirsten}
 \address{Department of Mathematics, Baylor University,
         Waco, TX 76798 U.S.A.}
\author{Paul Loya}
 \address{Department of Mathematics, Binghamton University,
Vestal Parkway East, Binghamton, NY 13902 U.S.A.}
 \author{Jinsung Park}\address{School of Mathematics, Korea Institute for Advanced
 Study 207-43, Cheongnyangni 2-dong, Dongdaemun-gu, Seoul 130-722 Korea}
 \begin{abstract}
In this contribution we announce a complete classification and new
exotic phenomena of the meromorphic structure of $\z$-functions
associated to conic manifolds proved in \cite{KLP1}. In
particular, we show that the meromorphic extensions of these
$\z$-functions have, in general, countably many logarithmic branch
cuts on the nonpositive real axis and unusual locations of poles
with arbitrarily large multiplicity. Moreover, we give a precise
algebraic-combinatorial formula to compute the coefficients of the
leading order terms of the singularities.
\end{abstract}
\section{Introduction}
It is well known, that a precise understanding of the meromorphic
structure of zeta functions for Laplace-type operators is very
important and its applications in many areas of mathematics and
physics are ubiquitous. For example, via its relation to the small
$t$-asymptotic expansion of the heat kernel, the zeta function
$\zeta (s, \Delta )$ associated with a Laplacian $\Delta$ on a
smooth manifold with or without boundary encodes geometrical and
topological information about the manifold, see e.g.\
\cite{BGiP95}. In some detail, we have for the scalar Laplacian
over a compact $n$-dimensional Riemannian manifold $M$ that
\begin{eqnarray*}
(4 \pi)^{\frac n2} \Gamma(s) \, \zeta(s , \Delta) \equiv
\frac{\mathrm{Vol}(M)}{s - \frac n2} \ \pm \ \frac{\sqrt{\pi}
\mathrm{Vol}(\partial M)}{2} \frac{1}{s - { \frac {n - 1}{2}}}
\end{eqnarray*} modulo a function that is analytic at $s = \frac
n2$, $s = { \frac {n-1}{2}}$, where the ``$+$" sign is used for
Neumann conditions, the ``$-$" sign is used for Dirichlet
conditions,
and $\mathrm{Vol}(M)$, respectively $\mathrm{Vol}(\partial M)$
denote as usual the volume of $M$, respectively $\partial M$.
Furthermore, it is known in the same context that the zeta
function $\zeta(s,\Delta)$ has a meromorphic extension to the
whole complex plane with at most simple poles at the points $s =
\frac{n - k}{2}\notin -\N_0$ for $k \in \mathbb{N}_0$ with $\N_0 =
\{0,1,2,\ldots\}$. Moreover, $\zeta(s,\Delta)$ is analytic at the
points $s \in - \N_0$. In particular, $\zeta (s,\Delta )$ is
analytic about $s=0$ which allows to define a zeta regularized
determinant. This has far reaching applications in quantum field
theory, see e.g.
\cite{dowk76-13-3224,gerald,BElE-etc94,HawS77,BKirK01} and in the
context of the Reidemeister-Franz torsion \cite{RS71}. There are
many other examples where the meromorphic structure of zeta
functions is crucial, found in, for instance, index theory, the
study of the Casimir effect, the evaluation of trace anamolies,
and so forth. We refer the reader to
\cite{byts96-266-1,BElE-etc94,BGiP95,BKirK01,vass03-388-279} for
reviews. The basic properties mentioned are valid for {\it smooth}
manifolds with local boundary conditions and are well known facts
that have been exploited for decades.

The aim of this contribution is to show that the properties for
zeta functions of Laplace-type operators on smooth manifolds are
very special indeed and so are the applications based upon this
structure. Here, we announce a new result for manifolds with
conical singularities whose zeta functions possess unusual
meromorphic structures unparalleled in the zeta function
literature for Laplacians. Thus, the usual structure totally
breaks down when the manifold has a conical singularity. We begin
in Section \ref{sec-conic} by reviewing the important subject of
conic manifolds introduced by Cheeger \cite{ChJ79,Ch} and which
appear in many areas of physics including when one studies the
Aharonov-Bohm potential \cite{AhY-BoD59} (see also
\cite{alfo89-62-1071,bene00-61-085019,sous89-124-229,hage90-64-503,lese98-193-317}),
classical solutions of Einstein's equations \cite{SoD-StA77},
cosmic strings \cite{ViA85}, global monopoles \cite{BaM-ViA89},
and the Rindler metric \cite{LCT18}, to name a few areas.
Afterwards, in Section \ref{sec-zetaconic}, we study the zeta
function associated to general self-adjoint extensions of
Laplace-type operators on conic manifolds and discuss their
extraordinary properties including countably many unusual poles
and logarithmic singularities. We also give an explicit
algebraic-combinatorial formula to compute these singularities and
show that such singularities occur even in simple examples.

Finally, we remark that one can always conjure
up``\emph{artificial}" zeta functions having unusual properties
compared to the ones described at the beginning. For example, the
zeta function associated with the prime numbers $P$, $$\zeta (s) =
\sum_{p\in P} p^{-s},$$ has a logarithmic branch cut at $s=1$, see
e.g. \cite{serre}. But for \emph{natural} zeta functions, that is
zeta functions of Laplacians on compact manifolds associated to
geometric or physical problems, the unusual properties described
here seem to be unique.

\section{Conic manifolds} \label{sec-conic}

In this section we study Laplacians on conic manifolds. One way to
understand operators over conic manifolds is to start with
simplest conic manifolds.

\subsection{Regions in $\R^2$ minus points} \label{ssec-egR2}

Let $\Omega \subset \R^2$ be any compact region and take polar
coordinates $(x,y) \longleftrightarrow (r,\theta)$ centered at any
fixed point in $\Omega$. In these coordinates, the metric takes
the form $dx^2 + dy^2 = dr^2 + r^2 \, d\theta^2$, which is called
a \emph{conic metric}. The standard Laplacian on $\R^2$ takes the
form
\[
\Delta_{\R^2} = - \partial_x^2 - \partial_y^2 = - \partial_r^2 -
\frac{1}{r} \partial_r - \frac{1}{r^2} \partial_\theta^2,
\]
and, finally, the measure transforms to $d x d y = r d r d
\theta$. Writing $\phi \in L^2(\Omega,r dr d \theta)$ as
\begin{equation} \label{phiiso}
\phi = r^{-1/2} \widetilde{\phi},
\end{equation}
where $\widetilde{\phi}:= r^{1/2} \phi$, we have
\[
\int_{\Omega} \phi(r,\theta)\, \psi(r,\theta)\, r d r d \theta =
\int_{\Omega} \widetilde{\phi}(r,\theta)\,
\widetilde{\psi}(r,\theta)\, d r d \theta .
\]
A short computation shows that
\[
\Delta_{\R^2} \phi = \left( - \partial_r^2 - \frac{1}{r}
\partial_r - \frac{1}{r^2} \partial_\theta^2 \right) \phi = r^{-1/2} \Delta\,
\widetilde{\phi},
\]
where $\Delta : = - \partial_r^2 + \frac{1}{r^{2}}
A_{\mathbb{S}^1}$ with $A_{\mathbb{S}^1} :=  -
\partial_\theta^2 - \frac14$.
In conclusion: Under the isomorphism (\ref{phiiso}) (called a
\emph{Liouville transformation}), $L^2(\Omega,r d r d \theta)$ is
identified with $L^2(\Omega, d r d \theta)$, and
\begin{equation} \label{DeltaR2}
\Delta_{\R^2} \quad  \longleftrightarrow \quad - \partial_r^2 +
\frac{1}{r^{2}} A_{\mathbb{S}^1},\qquad \text{where}\quad
A_{\mathbb{S}^1} = -
\partial_\theta^2 - \frac14 .
\end{equation}
Notice that the eigenvalues of $A_{\mathbb{S}^1}$ are given by
$\{k^2 - \frac14\, |\, k \in \Z\}$, in particular,
$A_{\mathbb{S}^1} \geq - \frac14$.

\subsection{Conic manifolds}
{ Let $M$ be a $n$-dimensional compact manifold with boundary
$\Gamma$ and let $g$ be a smooth Riemannian metric on $M \setminus
\partial M$. We assume that near $\Gamma$ there is a collared neighborhood
$\mathcal{U} \cong [0,\varepsilon)_r \times \Gamma$, where
$\varepsilon > 0$ and the metric $g$ is of product type $dr^2 +
r^2 h$ with $h$ a metric over $\Gamma$. Such a metric is called a
\emph{conic metric} and $M$ is called a \emph{conic manifold},
concepts introduced by Cheeger \cite{ChJ79,Ch}. As in the $\R^2$
case, using a Liouville transformation over the collar
$\mathcal{U}$, $L^2(M,dg)$ is identified with $L^2(M, dr dh)$ and
the scalar Laplacian $\Delta_g$ is identified with
\begin{equation} \label{Deltaconic}
\Delta_g \big|_{\mathcal{U}} = - \partial_r^2 +
\frac{1}{r^2}A_{\Gamma},\qquad \text{where}\quad A_{\Gamma} =
\Delta_\Gamma + \Big(\frac{1 - n}{2}\Big)
              \Big(1 + \frac{1 - n}{2}\Big)
\end{equation}
and $\Delta_\Gamma$ is the Laplacian over $\Gamma$. Notice that
$A_\Gamma \geq -\frac14$ because the function $x ( 1 + x)$ has the
minimum value $- \frac14$ (when $x = -\frac12$).}

Regular singular operators \cite{BS1} generalize the example
\eqref{Deltaconic} as follows. Let $E$ be a Hermitian vector
bundle over $M$, let $g$ be a metric on $M$ of product-type $g =
dr^2 + h$ over $\mathcal{U}$, and let $\Delta$ be a second order
elliptic differential operator over $M \setminus
\partial M$
that is symmetric on $C^\infty_c(M \setminus \partial M, E)$ such
that the restriction of $\Delta$ to $\mathcal{U}$ has the
``singular" form
\begin{equation} \label{D}
\Delta \big|_{\mathcal{U}} = - \partial_r^2 + \frac{1}{r^2}
A_\Gamma,
\end{equation}
where $A_\Gamma$ is a Laplace-type operator over $\Gamma$ with
$A_\Gamma \geq -\frac14$.\footnote{This condition is needed for
technical reasons; if $A_\Gamma \not\geq - \frac14$, then $\Delta$
is not bounded below \cite{BS1,CaC83}.} The operator $\Delta$ is
called a second order \emph{regular singular operator}. We remark
that the manifold $M$ may have boundary components up to which
$\Delta$ is smooth; at such components, we put local boundary
conditions such as the Dirichlet or Neumann boundary conditions
but we will not belabor this point. In view of \eqref{DeltaR2} and
\eqref{Deltaconic}, the Laplacian on a punctured region in $\R^2$
and the scalar Laplacian on a conic manifold are regular singular
operators. Other examples include the Laplacian on forms and
squares of Dirac operators on conic manifolds
\cite{BS1,ChJ79,Ch,ChA85,BKirK01,ULeMo,LMP}.

\subsection{Self-adjoint extensions}

In this kind of a setting there are different self-adjoint
extensions
\[
\Delta_{\mathfrak{D}} := \Delta : \mathfrak{D} \to L^2(M,E)
\]
possible, where $\mathfrak{D} \subset \mathfrak{D}_\max := \{\phi
\in L^2(M,E)\, |\, \Delta \phi \in L^2(M,E)\}$; for general
references on self-adjoint extensions of Laplacians and their
applications to physics see, e.g., \cite{albe88b,bonn01-69-322}.
From Von Neumann's theory of self-adjoint extensions
\cite{ChJ79,Ch,GM03,BLeM97,MoE99}, the self-adjoint extensions of
$\Delta$ are parameterized by Lagrangian subspaces in the
eigenspaces of $A_\Gamma$ with eigenvalues in the interval $[-
\frac14, \frac34)$. To describe these extensions, denote by
\begin{equation} \label{laq}
-\frac14 = \underbrace{\lambda_1 = \lambda_2 = \cdots =
\lambda_{q_0}}_{ = - \frac14} < \underbrace{\lambda_{q_0+1} \leq
\lambda_{q_0 + 2} \leq \cdots \leq \la_{q_0 + q_1}}_{- \frac14 <
\la_\ell < \frac34}
\end{equation}
the spectrum of $A_\Gamma$ in the finite interval $[- \frac14,
\frac34)$ where each eigenvalue is counted according to its
multiplicity.  Then the self-adjoint extensions of $\Delta$ are in
a one-to-one correspondence to the Lagrangian subspaces in
$\C^{2q}$ where $q = q_0 + q_1$. We note that (see, e.g.
\cite{KoV-ScR99}), a subspace $L \subset \C^{2q}$ is Lagrangian if
and only if there exists $q \times q$ matrices $\Aa$ and $\Bb$
such that the rank of the $q \times 2 q$ matrix $( \Aa\quad  \Bb )
$ is $q$, $\Aa' \, \Bb^*$ is self-adjoint where $\Aa'$ is the
matrix $\Aa$ with the first $q_0$ columns multiplied by $-1$, and
$L = \{ \phi \in \C^{2q} \, | \, ( \Aa  \quad \Bb)  \phi = 0\}$.
Given such a subspace $L \subset \C^{2q}$ there exists a
\emph{canonically} associated domain {$\mathfrak{D}_L \subset
\mathfrak{D}_\max$} such that $\Delta_L := \Delta : \mathfrak{D}_L
\to L^2(M,E)$ is self-adjoint. Any such self-adjoint extension has
a discrete spectrum \cite{BLeM97} and hence, if $\{\mu_j\}$
denotes the spectrum of $\Delta_L$, then we can form the
corresponding zeta function
\[
\zeta(s, \Delta_L) := \sum_{\mu_j \ne 0} \frac{1}{\mu_j^s}.
\]
For special self-adjoint extensions, like the Friedrichs
extension, the zeta function has been studied by many people going
back to the 70's
\cite{BKD,BS1,CaC83,CaC88,Ch,cogn94-49-1029,CoG-ZeS97,DowJ77,DowJ94,FurD94,BLeM97,LMP,SprM05};
the properties are similar to those for the smooth case described
in the Introduction except perhaps for an additional pole at $s =
0$. On the other hand, for general self-adjoint extensions, the
zeta function $\zeta(s, \Delta_L)$ has, in general, very
pathological properties that remained unobserved and that we shall
describe in the next section.

Zeta functions have also been studied for more general ``cone
operators,'' which generalize regular singular operators, see
e.g.\ Gil \cite{GilJ}. For recent and ongoing work involving
resolvents of general self-adjoint extensions of cone operators,
which is the first step to a full understanding of zeta functions,
see Gil et al.\ \cite{GKM1,GKM2} and Coriasco et al.\ \cite{CSS}.

\section{Pathological zeta functions on conic manifolds}
\label{sec-zetaconic}

In this section we state our theorem that completely classifies
the meromorphic structure of zeta functions $\zeta(s,\Delta_L)$
and we give concrete examples of the theorem.

\subsection{The main theorem}
Let $\Aa$ and $\Bb$ be $q \times q$ matrices defining a Lagrangian
$L \subset \C^{2q}$. Before stating the main result, we apply a
straightforward three-step algebraic-combinatorial algorithm to
$\Aa$ and $\Bb$ that we need for the statement.\\ {\bf Step 1:} We
define the function
\begin{eqnarray} \label{polyp}
p(x,y) := \det \left( \begin{array}{cc} \begin{array}{cc}
\hspace{1.0cm}\Aa &
\hspace{3.0cm}\Bb\end{array} \\
\begin{array}{cccc}
x\, \Id_{q_0} & 0 & 0 & 0\\
0 & \tau_1\, y^{2 \nu_1} & 0 & 0\\
0 & 0 & \ddots & 0\\
0 & 0 & 0 & \tau_{q_1}\, y^{2 \nu_{q_1}}
\end{array}
&  \Id_q
\end{array} \right),
\end{eqnarray}
where $\Id_k$ denotes the $k \times k$ identity matrix and where
\[
\nu_j := \sqrt{\la_{q_0 + j} + \frac14}\ , \quad \tau_j =
2^{2\nu_j} \frac{ \Gamma(1 + \nu_j)}{ \Gamma(1 - \nu_j)},\qquad j
= 1,\ldots, q_1.
\]
Here, $q_0,q_1,\la_j$ are explained in \eqref{laq}. Evaluating the
determinant, we can write $p(x,y)$ as a ``polynomial"
\[
p(x,y) = \sum a_{j \alpha} \, x^j\, y^{2 \alpha} ,
\]
where the $\alpha$'s are linear combinations of $\nu_1, \ldots,
\nu_{q_1}$ and the $a_{j \alpha}$'s are constants. Let $\alpha_0$
be the smallest of all $\alpha$'s with $a_{j \alpha} \ne 0$ and
let $j_0$ be the smallest of all $j$'s amongst the $a_{j \alpha_0}
\ne 0$. Then factoring out the term $a_{j_0 \alpha_0}\, x^{j_0}\,
y^{2 \alpha_0}$ in $p(x,y)$ we can write $p(x,y)$ in the form
\begin{equation} \label{pxy}
p(x,y) = a_{j_0 \alpha_0}\, x^{j_0}\, y^{2 \alpha_0} \Big( 1 +
\sum  b_{k \beta} \, x^k \, y^{2 \beta} \Big)
\end{equation}
for some constants $b_{k \beta}$ (equal to $a_{k \beta}/a_{j_0
\alpha_0}$).\\
{\bf Step 2:} Using formal power series expansion, we can write
\begin{equation} \label{clx}
\log \Big( 1 + \sum  b_{k \beta} x^k y^{2 \beta} \Big) = \sum
c_{\ell \xi} \, x^\ell \, y^{2 \xi}
\end{equation}
for some constants $c_{\ell \xi}$. The $\xi$'s appearing in
\eqref{clx} are nonnegative, countable, and approach $+\infty$
unless $\beta = 0$ is the only $\beta$ occurring in \eqref{pxy},
in which case only $\xi = 0$ occurs in \eqref{clx}. Also, the
$\ell$'s with $c_{\ell \xi} \ne 0$ for a fixed $\xi$ are bounded
below.\\
{\bf Step 3:} For each $\xi$ appearing in \eqref{clx}, define
\begin{equation} \label{ellxi}
p_\xi := \min \{\ell \leq 0 \, |\, c_{\ell \xi} \ne 0\}\qquad
\text{and}\qquad \ell_\xi := \min \{\ell > 0 \, |\, c_{\ell \xi}
\ne 0\},
\end{equation}
whenever the sets $\{\ell \leq 0 \, |\, c_{\ell \xi} \ne 0\}$ and
$\{\ell > 0 \, |\, c_{\ell \xi} \ne 0\}$, respectively, are
nonempty. Let {${\mathscr P}$}, respectively {${\mathscr L}$},
denote the set of $\xi$ values for which the respective sets are
nonempty. The following theorem is our main result \cite{KLP1}.\\
{\bf Theorem 3.1:} \label{thm-main} For an arbitrary Lagrangian
$L$, the $\zeta$-function $\zeta(s,\Delta_L)$ extends from $\Re s
> \frac{n}{2}$ to a holomorphic function on $\C \setminus
(-\infty,0]$. Moreover, $\zeta(s,\Delta_L)$ can be written in the
form
\[
\zeta(s,\Delta_L) = \zeta_{\mathrm{reg}}(s,\Delta_L) +
\zeta_{\mathrm{sing}}(s,\Delta_L),
\]
where $\zeta_{\mathrm{reg}}(s,\Delta_L)$ has possible simple poles
at the usual locations $s = \frac{n - k}{2}$ with $s \notin -\N_0$
for $k \in \N_0$ and at $s = 0$ if $\dim \Gamma > 0$, and where
$\zeta_{\mathrm{sing}}(s,\Delta_L)$ has the following expansion:
\begin{eqnarray} \label{zetasing}
\zeta_{\mathrm{sing}}(s,\Delta_L)& =& \frac{\sin (\pi s)}{\pi}
\bigg\{ (j_0 - q_0) e^{-2 s (\log 2 - \gamma)} \log s\\ & &\quad
\quad + \sum_{ \xi \in {\mathscr P}} \frac{f_\xi(s)}{(s +
\xi)^{|p_\xi| + 1}} + \sum_{ \xi \in {\mathscr L}} g_\xi(s) \log
(s + \xi) \bigg\},\nonumber
\end{eqnarray}
where $j_0$ appears in \eqref{pxy} and $f_\xi(s)$ and $g_\xi(s)$
are entire functions of $s$ such that
\[
f_\xi(-\xi) = (-1)^{|p_\xi|+1} c_{p_\xi \xi} \,
\frac{|p_\xi|!}{2^{|p_\xi|}} \, \xi
\]
and
\begin{eqnarray}
g_\xi(s) = \left\{\begin{array}{ll} c_{\ell_0, 0} \,
\frac{2^{\ell_0}}{ (\ell_0 - 1)!} s^{\ell_0} +
\mathcal{O}(s^{\ell_0 + 1}) & \text{if
$\xi = 0$,}\\[.1cm] - c_{\ell_\xi \xi} \, \frac{\xi\,\,
2^{\ell_\xi}}{(\ell_\xi - 1)!} (s + \xi)^{\ell_\xi - 1} +
\mathcal{O}((s + \xi)^{\ell_\xi}) & \text{if $\xi > 0$.}
\end{array}\right.
\end{eqnarray}
{\bf Remark 3.2:} This theorem is very simple to use in practice
and gives precise results \emph{immediately} as we show in the
following subsection. The regular part
$\zeta_{\mathrm{reg}}(s,\Delta_L)$ will only have possible poles
at $s = \frac{n}{2} - k \notin -\N_0$ in the case that $\Gamma$ is
the only boundary component of $M$ and the residue of
$\zeta_{\mathrm{reg}}(s,\Delta_L)$ at $s = 0$ is given by
\[
\mathrm{Res}_{s = 0} \zeta_{\mathrm{reg}}(s,\Delta_L) = - \frac12
\mathrm{Res}_{s = - \frac12} \zeta (s, A_\Gamma);
\]
in particular, this vanishes if $\zeta (s, A_\Gamma )$ is in fact
analytic at $s = -\frac12$. The expansion \eqref{zetasing} means
that for any $N \in \N$,
\begin{eqnarray*}
\zeta_{\mathrm{sing}}(s,\Delta_L) &=& \frac{\sin (\pi s)}{\pi}
\bigg\{ (j_0 - q_0) e^{-2 s (\log 2 - \gamma)} \log s  + \sum_{
\xi \in {\mathscr P} ,\, \xi \leq N} \frac{f_\xi(s)}{(s +
\xi)^{|p_\xi| + 1}} \\& &\quad \quad + \sum_{ \xi\in {\mathscr L}
,\, \xi \leq N} g_\xi(s) \log (s + \xi) \bigg\} + F_N(s),
\end{eqnarray*}
where $F_N(s)$ is holomorphic for $\Re s \geq - N$. Finally, for
arbitrary self-adjoint extensions with $A_\Gamma \geq - \frac14$,
the $\zeta$-function has been studied by Falomir, Muschietti and
Pisani \cite{FMP} (see also \cite{FPW} and joint work with Seeley
\cite{FMPS}) for one-dimensional Laplace-type operators over
$[0,1]$ and by Mooers \cite{MoE99} who studied the general case of
operators over manifolds and who was the first to notice the
presence of unusual poles. \\
{\bf Remark 3.3:} There are equally pathological heat operator and
resolvent trace expansions with exotic behaviors such as
logarithmic terms of arbitrary positive and negative multiplicity;
we refer the reader to \cite{KLP1} for the details.

\subsection{Examples of Theorem \ref{thm-main}}

{\bf Example 1:} Falomir \emph{et al.}\ \cite{FMP} study the
operator
\[
\Delta = - \frac{d^2}{dr^2} + \frac{1}{r^2}\, \la\qquad
\text{over}\ \ [0,1]
\]
with the Dirichlet or Neumann condition at $r=1$ and $- \frac14
\leq \la < \frac34$; thus, in this example, ``$A_\Gamma$" is the
number ``$\la$." In this case, $V = \C^2$, therefore Lagrangians
$L \subset \C^2$ are determined by $1 \times 1$ matrices (numbers)
$\Aa = \alpha$ and $\Bb = \beta$, not both zero, such that $\alpha
\overline{\beta}$ is real. Fix such an $(\alpha,\beta)$ and let us
assume that $- \frac14 < \la < \frac34$ so there is no $- \frac14$
eigenvalue (we will come back to the $\la = - \frac14$ in a
moment). Then with $\nu := \sqrt{\la + \frac14}$ and $\tau :=
2^{2\nu} \frac{ \Gamma(1 + \nu)}{ \Gamma(1 - \nu)}$,
\[
p(x,y) := \det
\left(\begin{array}{cc} \alpha & \beta \\
\tau \, y^{2 \nu} & 1 \end{array}\right) = \alpha - \beta \,
\tau\, y^{2 \nu} = \alpha\Big( 1 - \frac{\tau \beta}{\alpha} y^{2
\nu} \Big),
\]
where we assume that $\alpha,\beta \ne 0$ (the $\alpha = 0$ or
$\beta = 0$ cases can be handled easily), and we write $p(x,y)$ as
\eqref{pxy}. Forming the power series \eqref{clx}, we see that
\[
\log \Big( 1 - \frac{\tau \beta}{\alpha} y^{2 \nu} \Big) = \sum_{k
= 1}^\infty \frac{(-1)^{k-1}}{k} \Big(- \frac{\tau \beta}{\alpha}
y^{2 \nu} \Big)^k = \sum_{k = 1}^\infty c_{0, \nu k}\, x^0 y^{2
\nu k},
\]
where $c_{0, \nu k} = - \frac{1}{k} \Big( \frac{\tau
\beta}{\alpha} \Big)^k$ and where the $\xi$'s in \eqref{clx} are
given by the $\nu k$'s and the $\ell$'s in \eqref{clx} are all
$0$. Using the definition \eqref{ellxi} for $p_\xi$ and
$\ell_{\xi}$, we immediately see that $\ell_{\nu k}$ is never
defined, while
\[
p_{\nu k} = \min \{\ell \leq 0\, |\, c_{\ell , \nu k} \ne 0\} = 0
\]
exists for all $k \in \N$. Therefore, by Theorem \ref{thm-main},
\[
\zeta_{\mathrm{sing}}(s,\Delta_L) = \frac{\sin (\pi s)}{\pi}
\sum_{k = 1}^\infty \frac{f_k(s)}{s + \nu k}
\]
with $f_k(s)$ an entire function of $s$ such that
\[
f_k(-\nu k) = -  c_{0, \nu k} \frac{0!}{2^0} \nu k = \nu \Big(
\frac{\tau \beta}{\alpha} \Big)^k .
\]
In particular, $\zeta_{\mathrm{sing}}(s,\Delta_L)$ has possible
poles at each $s = - \nu k$ with residue equal to
\[
\mathrm{Res}_{s = - \nu k} \zeta_{\mathrm{sing}}(s ,\Delta_L) =
\frac{\sin (\pi (-\nu k))}{\pi} \nu \Big( \frac{\tau
\beta}{\alpha} \Big)^k = - \frac{\nu \sin \pi \nu k}{\pi} \Big(
\frac{\tau \beta}{\alpha} \Big)^k,
\]
which is the main result of \cite{FMP} (see Equation (7.11) of
loc.\ cit.).

Assume now that $\la = - \frac14$. In this case,
\[
p(x,y) := \det
\left(\begin{array}{cc} \alpha & \beta \\
x & 1 \end{array} \right) = \alpha - \beta \, x = \alpha\Big( 1 -
\frac{\beta}{\alpha} x \Big),
\]
where we assume that $\alpha,\beta \ne 0$ (the $\alpha = 0$ or
$\beta = 0$ cases can be handled easily).
Proceeding as before, by Theorem \ref{thm-main},
\[
\zeta_{\mathrm{sing}}(s,\Delta_L) = \frac{\sin (\pi s)}{\pi}
\bigg\{ - e^{-2 s (\log 2 - \gamma)} \log s + g_0(s) \log s
\bigg\},
\]
$g_0(s)$ being an entire function of $s$ such that $g_0(s) =
\mathcal{O}(s)$. In particular, $\zeta(s,\Delta_L)$ has a
\emph{genuine} logarithmic singularity at $s = 0$.  When $\beta =
0$, one can easily check that we still have a logarithmic
singularity at $s = 0$ and when $\alpha = 0$, we only have the
part $\zeta_{\mathrm{reg}}(s,\Delta_L)$ and no
$\zeta_{\mathrm{sing}}(s,\Delta_L)$; one can easily show that (see
\cite{BS1}) $\alpha = 0$ corresponds to the Friedrichs extension;
thus we can see that $\zeta(s,\Delta_L)$ has a logarithmic
singularity for all
extensions except the Friedrichs.\\
{\bf Example 2:} \label{eg-Lap} (The Laplacian on $\R^2$) If
$\Delta$ is the Laplacian on a compact region in $\R^2$, then as
we saw before in Section \ref{ssec-egR2}, $A_\Gamma$ has a
$-\frac14$ eigenvalue of multiplicity one and no eigenvalues in
$(- \frac14, \frac34)$. Therefore, the exact same argument we used
in the $\la = - \frac14$ case of the previous example shows that
$\zeta(s,\Delta_L)$ has a logarithmic singularity for all
extensions except the Friedrichs.\\
{\bf Example 3:} \label{eg-countk} Consider now the case of a
regular singular operator $\Delta$ over a compact manifold and
suppose that $A_\Gamma$ has two eigenvalues in
$[-\frac14,\frac34)$, the eigenvalue $- \frac14$ and another
eigenvalue $- \frac14 < \la < \frac34$, both of multiplicity one.
This situation occurs, for example, in the two-dimensional flat
cone in $\R^3$ with $\Gamma = \mathbb{S}^1_\nu$ where
$\mathbb{S}^1_\nu$ is the circle with metric $d\theta/\nu$ where
$\frac12 < \nu < 1$; indeed, after a Liouville transformation, we
have
\[
A_\Gamma = - \nu^2 \partial_\theta^2 - \frac14 ,
\]
which only has the eigenvalues $-\frac14$ and $\la = \nu^2 -
\frac14$ in the interval $[-\frac14,\frac34)$. In this case, $q =
2$ and Lagrangians $L \subset \C^4$ are determined by $2 \times 2$
matrices $\Aa$ and $\Bb$ such that $( \Aa \quad \Bb )$ has full
rank and $\Aa ' \Bb^*$ is self-adjoint. Consider the specific
examples
\[
\Aa = \left(\begin{array}{cc} 0 & 1 \\
-1 & 0
\end{array}\right) \ , \quad  \Bb = \Id  .
\]
Then with $\tau := 2^{2\nu} \frac{ \Gamma(1 + \nu)}{ \Gamma(1 -
\nu)}$ where $\nu = \sqrt{\la + \frac14}$, we have
\[
p(x,y) := \det
\left(\begin{array}{cccc} 0 & 1 & 1 & 0 \\
-1 & 0 & 0 & 1 \\ x & 0 & 1 & 0 \\ 0 & \tau \, y^{2 \nu} & 0 & 1
\end{array}\right) = 1 + \tau\, x \, y^{2 \nu} .
\]
Forming the power series \eqref{clx}, we see that
\[
\log \Big( 1 + \tau\, x \, y^{2 \nu} \Big) = \sum_{k = 1}^\infty
\frac{(-1)^{k-1}}{k} \Big( \tau\, x \, y^{2 \nu} \Big)^k = \sum_{k
= 1}^\infty c_{k, \nu k}\, x^k y^{2 \nu k},
\]
where $c_{k, \nu k} = (-1)^{k-1} \frac{\tau^k}{k}$. Using the
definition \eqref{ellxi} for $p_\xi$ and $\ell_{\xi}$,  we
immediately see that $p_{\nu k}$ is never defined, while each
$\ell_{\nu k}$ is defined:
\[
\ell_{\nu k} = \min \{\ell > 0\, |\, c_{\ell , \nu k} \ne 0\} = k
.
\]
Therefore, by Theorem \ref{thm-main},
\[
\zeta_{\mathrm{sing}}(s,\Delta_L) = \frac{\sin (\pi s)}{\pi}
\bigg\{ - e^{-2 s (\log 2 - \gamma)} \log s + \sum_{k = 1}^\infty
g_k(s) \log (s + \nu k) \bigg\},
\]
with $g_k(s)$ an entire function of $s$ such that
\[
g_k(s) = (-1)^k \frac{\tau^k 2^{k} \nu}{(k - 1)!} (s + \nu k)^{k -
1} + \mathcal{O}((s + \nu k)^{k}) .
\]
In particular, $\zeta_{\mathrm{sing}}(s,\Delta_L)$ has
\emph{countably} many logarithmic singularities!\\
{\bf Example 4:} With the same situation as in the previous
example, consider
\[
\Aa = \left(\begin{array}{cc} - 1 & 1 \\
0 & 0 \end{array}\right) \ , \quad \Bb = \left(\begin{array}{cc} 0 & 0 \\
1 & -1
\end{array}\right) ,
\]
so that
\[
p(x,y) := \det
\left(\begin{array}{cccc} -1 & 1 & 0 & 0 \\
0 & 0 & 1 & -1 \\ x & 0 & 1 & 0 \\ 0 & \tau \, y^{2 \nu} & 0 & 1
\end{array}\right) = x - \tau\, y^{2 \nu} = x \Big(1 - \tau \, x^{-1} y^{2 \nu}
\Big).
\]
Proceedings as before, by Theorem \ref{thm-main},
\[
\zeta_{\mathrm{sing}}(s,\Delta_L) = \frac{\sin (\pi s)}{\pi}
\sum_{k = 1}^\infty \frac{f_k(s)}{(s + \nu k)^{k + 1}} ,
\]
with $f_k(s)$ an entire function of $s$ such that
\[
f_k(-\nu k) = (-1)^{k+1} c_{-k, \nu k} \, \frac{|-k|!}{2^{|-k|}}
\, \nu k = (-1)^k \frac{\tau^k k! \nu }{2^{k}} .
\]
In particular,  $\zeta_{\mathrm{sing}}(s,\Delta_L)$ has poles of
\emph{arbitrarily} large order!\\
{\bf Example 5:}
\label{eg-count3k} Consider now the case of a regular singular
operator $\Delta$ over a compact manifold such that $A_\Gamma$ has
three eigenvalues in $[-\frac14,\frac34)$, the eigenvalue $-
\frac14$ with multiplicity two and another eigenvalue $- \frac14 <
\la < \frac34$ of multiplicity one. This situation occurs, for
example, in the two-dimensional flat cone in $\R^3$ with $\Gamma =
\mathbb{S}^1 \sqcup \mathbb{S}^1_\nu$, the disjoint union of the
standard circle with metric $d \theta$ and the circle with metric
$d\theta/\nu$ where $\frac12 < \nu < 1$; indeed, after a Liouville
transformation, we have
\[
A_\Gamma = \Big( - \partial_\theta^2 - \frac14 \Big) \oplus \Big(
- \nu^2 \partial_\theta^2 - \frac14 \Big),
\]
where in the interval $[-\frac14,\frac34)$, the first operator has
only the $-\frac14$ eigenvalue and the second operator has only
the eigenvalues $-\frac14$ and $\la = \nu^2 - \frac14$. In this
case, $V = \C^6$ and Lagrangians $L \subset \C^6$ are determined
by $3 \times 3$ matrices $\Aa$ and $\Bb$. Consider the specific
examples
\[
\Aa = \left(\begin{array}{ccc} 0 & 1 & -1 \\
1 & 0 & 0 \\
1 & 0 & 0
\end{array}\right) \ , \quad \Bb = \Id.
\]
Then with $\nu = \sqrt{\la + \frac14}$ and $\tau := 2^{2\nu}
\frac{ \Gamma(1 + \nu)}{ \Gamma(1 - \nu)}$, using the procedure
outlined several times, we find
\begin{eqnarray*}
\zeta_{\mathrm{sing}}(s,\Delta_L) &=& \frac{\sin (\pi s)}{\pi}
\bigg\{ - e^{-2 s (\log 2 - \gamma)} \log s + \sum_{k = 1}^\infty
\frac{f_k(s)}{(s + \nu k )^{k + 1}}\\
& &\quad \quad \quad \quad + \sum_{k = 1}^\infty g_k(s) \log (s +
\nu k) \bigg\},
\end{eqnarray*}
where $f_k(s)$ and $g_k(s)$ are entire functions of $s$ such that
\[
f_k(-\nu k) = (-1)^{k+1} c_{-k, \nu k } \, \frac{k!}{2^{k}} \, \nu
k = \frac{(-1)^k}{k} \tau^k \, \frac{k!}{2^{k}}\, \nu k = \,
(-1)^k \frac{ \tau^k k! \nu }{2^{k}}
\]
and
\begin{eqnarray*}
\hspace{-2.50cm}g_k(s) = 2 \nu (-1)^{m+1} \tau^k {k \choose m+1}
\times\left\{
\begin{array}{ll}
1 + \mathcal{O}((s + \nu k)) & \text{if $k = 2 m +
1$ is odd,}\\
2 (s + \nu k) + \mathcal{O}((s + \nu k)^{2}) & \text{if $k = 2 m$
is even.}
\end{array}\right.
\end{eqnarray*}
In particular, $\zeta_{\mathrm{sing}}(s,\Delta_L)$ has poles of
\emph{arbitrarily} high orders and in addition to a logarithmic
singularity at the origin, \emph{countably} many logarithmic
singularities at the same locations of the poles!\\
{\bf Example 6:} From the previous examples, we can see that by
looking at flat cones in $\R^3$ whose boundaries are disjoint
unions of circles of various circumferences, one can easily come
up with completely natural (that is, geometric) zeta functions
having as wild singularities involving unusual poles and
logarithmic singularities as the mind can image.

\subsection{Conclusion and final remarks}

In this paper we have considered zeta functions of self-adjoint
extensions of Laplace-type operators over conic manifolds. We have
presented a theorem that gives the exact structure of zeta
functions for arbitrary self-adjoint extensions of Laplace-type
operators over manifolds with conical singularities. As we have
seen, the structure found can be dramatically different from the
standard one. Using this exact structure, with a suitable
redefinition, functional determinants of Laplacians on generalized
cones can still be obtained \cite{KLP2}.

The ideas presented here can equally well be applied to the Dirac
operator {\cite{KLP3}}. In the presence of a Dirac delta magnetic
field \cite{manu93-301-72}, different self-adjoint extensions are
considered as manifestations of different physics within the
vortex \cite{sous89-40-1346}. The physics represented by the
self-adjoint extensions described by ${\mathcal A}$ and ${\mathcal
B}$ and the implications of the meromorphic structure of the zeta
functions found are very interesting questions to pursue.

\section*{Acknowledgements}
KK acknowledges support by the Baylor University Summer Sabbatical
Program and by the Baylor University Research Committee.
 
 \end{document}